# Gender gap in mobility outside home in urban India

Rahul Goel
Transportation Research and Injury Prevention Centre, Indian Institute of Technology Delhi, New Delhi, India; rgoel@iitd.ac.in; rahulatiitd@gmail.com

**ABSTRACT**
India has one of the highest levels of gender inequality in the world. Work participation rate of women is among the lowest with a wide gender gap. There are seclusion norms that restrict mobility of women outside home. However, transport literature in India has not explored the impact of this lack of autonomy on gender differences in travel demand. I use 2019 population-representative nationwide time-use survey of India. The dataset reported both travel and non-travel activities in 30-minute episodes over a 24-hour period. For urban residents, I analysed gender differences in trip rates and mobility rates, where the latter is defined as the percentage going out of home at least once on the reporting day. I developed gender-stratified logistic regression models at the individual level with mobility as a binary outcome. It was found that 53% of the females did not report going out of home compared to only 14% males. The mobility of females reduces steeply from adolescence to young adulthood and then remains largely stable at a low level before reducing further for older adults. No such variation is observed among males, except their mobility is also reduced among older adults. There is a clear dichotomy with women mostly participating in in-house activities while men mostly involved in out-of-home activities. Adolescence or adulthood, marriage, living with one or more household members, having an infant in the house, lower income, and less education are associated with lower likelihood of female mobility. I discuss many implications of these gender differences in mobility.

## 1      INTRODUCTION

India has one of the highest levels of gender inequality in the world. In 2021, World Economic Forum ranked India at 140 out of 156 countries for which they reported Gender Gap Index (WEF, 2021). The index included following four components— economic participation and opportunity, educational attainment, health and survival, and political empowerment. The countries that ranked lower than India included those from South Asia (e.g. Pakistan), Middle East (e.g. Oman), and North Africa (e.g. Morocco). Among the four components of the index, India fares the worst in economic participation. Labour force participation rate of women in India is among the lowest in the world with high gender gap— in urban areas, the rate is 20% among women compared to 75% among men in 2017 (Deshpande, 2021; Deshpande & Kabeer, 2019).

One important indicator of gender inequality in India is the lack of autonomy among women to go outside home resulting from seclusion norms. These norms constraint women's freedom of movement outside home (Kantor, 2002). According to a 2011-12 population-representative nationwide survey (IHDS-II, 2022), 78% women in urban areas reported that they need "permission" to visit a grocery store, 89% need it to visit a relative or a friend in the neighbourhood, and 94% to travel for a short distance by train or bus. The permission is often reported to be asked from a male member in the household. To some extent, gender inequality in economic participation and educational attainment can be attributed to these seclusion norms. If women cannot go out, it limits their options to engage in activities.

As expected, limited movement of women outside home is reflected in their travel patterns. According to Census of India conducted in 2011, among all the workers that reported travelling to work "outside home", only 20% of those were women. The underrepresentation of females is not limited to work trips alone. Using GPS to measure daily movement of individuals in a peri-urban area in India, (Sanchez et al., 2017) reported that men make 3 times as many trips per day (for all purpose) as women. Using data from a household travel survey, (Goel et al., 2022) also reported greater level of immobility among women than men in Delhi. Immobility is the percentage of individuals reporting not going out of home (or making any trip) on the survey day (Madre et al., 2007). For Delhi, 58% of females reported not going out of house on the survey day, while for males this was 26% percentage point lower. For other cities in India, (Mahadevia & Advani, 2016) and (Saigal et al., 2020) also reported similarly high levels of immobility among women and large gender gap in Rajkot and Jaipur, respectively. One contradictory finding is by (Srinivasan & Rogers, 2005), who reported greater number of trips made by females than males for an Indian city, though the study was limited to a small sample.

The above discussion strongly indicates that there is high level of gender inequality in India in terms of mobility outside home. This is reflected in the number of trips that individuals make in a day (i.e., per capita trip rate) or the percentage population that report going out of home at least once on the survey day. Despite such alarmingly low levels of autonomy among females, transport literature in India has not given this issue the attention it deserves. In any case, transport literature that investigated gender differences in travel behaviour is itself limited in India. In this paper, I will investigate this gender difference in mobility and its determinants.

There are many studies in India that analyse travel behaviour, however, they account for gender differences by specifying it as a binary variable in regression modelling, thus treating it as a confounder (Arasan et al., 1998). Such an approach (wrongly) assumes that all other variables impact travel behaviour equally for both genders, and that gender is not an effect modifier. This limited view of gender contributes little to our understanding of the gendering process i.e. how certain activities and responsibilities are treated socially appropriate for females. For instance, having children at home may result in different constraints for women than men, as the former may be responsible for accompanying them to school and for other care work. Much of this literature also investigates trip characteristics such as purpose, distance, and mode choice, where the latter is an important step in four-step transport modelling process (Jain & Tiwari, 2020; Manoj & Verma, 2015). In other words, it does not analyse whether an individual makes a trip or not, rather it looks at the differences in the characteristics of the reported trips.

Additionally, none of these studies aimed at understanding gender differences as their primary outcome, with only few exceptions. Alberts et al. (2016) used a qualitative approach to study livelihood strategies of women living in the periphery of a metropolitan city in India. The authors found childcare and household responsibilities, cultural norms prohibiting work outside home, and poor accessibility of jobs among the reasons for not working outside home. The authors concluded that the provision of public transportation contributes to livelihoods. Interestingly, most respondents did not mention distance, cost, or burden of traveling as a reason to stay at home. For low-income settlement in Delhi, Anand & Tiwari, (2006) reported greater levels of transport poverty among women resulting from dependence on slower modes of transport (walking and public transport) compared to men who also use cycles and motorcycles. This results from social norms that restrict use of certain modes and weaker access to household resources. Their impaired mobility is reflected in their work participation which remains low because of their limited access to jobs in the city.

Given the abysmal rate of work participation among women in India, there has generally been a lot of research focus on this issue (Deshpande & Singh, 2021). Paid labour also leads to more autonomy than unpaid household labour (Jayachandran, 2021). There is large literature from development economics in India that studies women empowerment and its many determinants. The ability to go outside home is treated as an indicator of women's autonomy in intra-household decisions (Choudhuri & Desai, 2020). With greater autonomy to travel outside, women are more likely to pursue education, work outside home and develop a social network. Since work participation rate and women's autonomy are both related to their mobility outside home, I will briefly review literature that has studied these two dimensions of women's empowerment.

Majority of Indian households are patrilocal in nature, which is the practice of married couple residing in or near husband's parents. In this context, intergenerational power dynamics can determine a married woman's autonomy. Khalil & Mookerjee (2019) reported that married women in patrilocal households are less likely to have freedom of movement or participate in economic decisions. Anukriti et al. (2020) also reported for a rural region in India that co-residence of a woman with her mother-in-law is strongly negatively associated with her mobility and her ability to form close social connections outside the home. Mehta & Sai (2021) reported that this lack of autonomy for mobility outside home is greater among daughters-in-law than the spouses or daughters of the head of household. Dhanaraj & Mahambare (2019) reported that living in a joint family reduces the chances of women working in a non-farm employment. Chakraborty et al. (2018) reported a similar impact on work participation of women due to greater household size.

Sudarshan & Bhattacharya (2009) also found labour force participation of married women to be significantly lower than those of unmarried women. They found that married women believed that they should stop working when their children are young but supported the idea of women re-joining the workforce when their children reach a certain age. The three key factors associated with greater work participation of women are higher education, reduction in time spent on housework (domestic technology, water and electricity, childcare arrangements), and safety in public spaces (transport, lighting). Similar findings were reported by Sarkar et al. (2019) using two rounds of a panel survey. The authors found greater probability of exit of women from workforce resulting from having a new-born child. The effect was opposite for those with older children, possibly resulting from reduction in childcare needs. However, Deshpande & Singh (2021) have recently argued that the low levels of work participation among women in India may not be due to their demographic change, for example, child-birth. Rather, it is the lack of steady employment that is keeping women out of workforce. Using a longitudinal dataset, the authors reported that the percentage of women who were in the work force at least once is much greater than the average rate of women in workforce.

The above discussion highlights the impact of various demographic, socioeconomic, and sociocultural variables on women's likelihood to work outside home. I will use this understanding to investigate a broader issue of gender differences in mobility outside home, and not just the mobility related to work. The literature on work participation limits its focus on working age group, and therefore, gender differences among children and older adults remain underrepresented in such work. In this study, I will investigate gender differences for all age groups. I present analysis of a nationwide time-use survey which records all activities within a 24-hour period. I restrict my analysis to urban residents. Household activities are often unequally distributed among the household members, and in Indian context, the burden falls mostly on women. Therefore, I will analyse mobility not only with respect to demographics and socioeconomic characteristics but also participation of individuals in other non-travel activities. The two outcomes of interest are the trip rate which is the average number of trips made in a day and the mobility rate which we define as the percentage of population that reported going out of home at least once on the reporting day.

## 2     DATA

I use "Time Use in India-2019" dataset, a population-representative time-use survey of a sample of households in India (MOSPI, 2020). The survey was conducted by National Statistical Office from January to December 2019 in four rounds of 3-month period each, thus representing all seasonal variations, and covered 82,897 households in rural areas and 55,902 households in urban areas. The questionnaire was administered to all members of the households of 6 years of age or above, using face-to-face interviews of respondents. In the absence of a respondent at the time of survey, another household member or someone outside the household was interviewed as proxy.

The survey recorded activities in 30-minute resolution from 4:00 AM of the day prior to the survey day to the 4:00 AM of the survey day. For each activity, there is an activity type, the type of location where the activity has been conducted (household dwelling, outside household, or non-fixed), and whether there is a simultaneous activity being performed. The activities are classified into one of 164 activity types under the following 9 major divisions— i) employment and related activities, ii) production of goods for own use, iii) unpaid domestic services for household members, iv) unpaid caregiving services for household members, v) unpaid volunteer, trainee and other unpaid work, vi) learning, vii) socializing and communication, community participation and religious practice, viii) culture, leisure, mass-media and sports practices and ix) self-care and maintenance. All activities of at least 10 minutes of duration within a given 30-minute slot were recorded, with a maximum of 3 activities within the same slot, along with a question asking if the activities were conducted simultaneously or not. The survey included individual characteristics of all household members. This includes their activity status reported under 14 categories that are broadly divided into 3 sub-categories of working, unemployed and not in labour force.

i) **working (or employed) category include the following:** *self-employed:* 1) worked in household enterprises (self-employed) as own-account worker 2) worked in household enterprises (self-employed) as an employer 3) worked in household enterprises (self-employed) as helper *regular wage/ salaried employee:* 4) worked as regular wage/salaried employee *casual labour:* 5) worked as casual labour in public works, 6) worked as casual labour in other types of works

ii) **not working but seeking/available for work (or unemployed) include the following:** 7) sought work or did not seek but was available for work

iii) **neither working nor available for work (or not in labour force) include the following:** 8) attended educational institutions, 9) attended to domestic duties only, 10) attended to domestic duties and was also engaged in free collection of goods (vegetables, roots, firewood, cattle feed, etc.), sewing, tailoring, weaving, etc. for household use, 11) rentiers, pensioners, remittance recipients, etc., 12) not able to work owing to disability, 13) others (including beggars, prostitutes, etc.), and 14) children of age 0-4 years.

The classification into one of these 14 categories is based on usual principal status approach according to which activity status of a person is that on which a person spent relatively long time (major time criterion) during the 365 days preceding the date of survey. The industry in which the person is engaged are based on National Industrial Classification (NIC) 2008 (MOSPI, 2008). We only used broad classification of different industries into 21 sections such as agriculture, mining, and manufacturing.

## 3 METHOD
### 3.1 Extraction of travel behaviour from time use survey

For this analysis, I selected only individuals living in urban areas, that includes 84207 females and 88914 males with a minimum age of 6 years. Among the reported activities, some activities are classified as travel activities, for example, going to work. However, there are several activities that are reported to occur outside household premises but did not have any travel activities accompanying those. All of these are listed in the **Supplementary Information (SI) Table S1**. I assume that it is because, for those trips, travel component may be too short (<10 minutes) to be considered as an activity on its own. The impact of low temporal resolution of time-use surveys on poor capturing of short-duration trips has also been discussed in the literature (Gerike et al., 2015). For example, a 30-minute shopping activity with no travel preceding or following it could be 20-min shopping activity with 5 minutes of travel on either end. For these activities, we imputed two trips if the person was reported to be at home before as well as after that reported out-of-home activity. If those activities were taking place consecutive to another reported travel activity, we then assigned only one trip, completing a tour. For example, if a shopping activity is reported after a return trip from work, it was assumed that the person made a trip from work to shopping location and from there returned home, and therefore, making two instead of one trip in her return trip. For these imputed trips, however, the duration of travel activities is not known.

### 3.2 Predictors of mobility

I present descriptive statistics of the sample stratified by females and males, and for the two groups combined. To conduct exploratory analysis, I fit bivariate generalised additive models (GAM) at individual level with number of daily trips as outcome, and individual and household characteristics as explanatory variables. GAM are generalised linear models in which outcome is linearly related to a smooth function of the predictor and I used "mgcv" library in R statistical software to develop these. I used graphical representation of these model outputs to understand the relationships. Next, I developed logistic regression models at the individual level. I used a binary outcome, with a value of one for those individuals who made at least one trip during the survey day and a value of zero otherwise. For a gender-stratified analysis, we developed regression models specific to females and males. I used individual and household demographics and socioeconomic characteristics as explanatory variables. In addition, we used time spent in different activity divisions as explanatory variables. I compared the average time spent in different activity divisions between males and females for different age groups (see **SI Table S2**). The gender differences are the greatest in the participation of employment, domestic, and caregiving activities. I expressed the three activity types as categorical variables with categories defined by the hours of time spent. I expressed monthly household expenditure as monthly per-capita expenditure (MPCE) and further standardised the variable. Two sets of models were developed. In one set, we included all the individual and household-level demographic and socioeconomic characteristics. In the second set, we also controlled for the participation time of different activity types. I also investigated the gender differences in the types of employment and how they relate to the work-related mobility. Next, I present gender differences in trip purpose and its variation with age.

## 4   RESULTS

**Table 1** presents descriptive statistics of the time-use survey respondents (6 years or older) stratified by two gender groups. Also presented in the table are the mobility rate and daily trip rate for each category and classified by gender.

There is a wide gender gap in terms of work participation. Only 16% females are either engaged in household enterprises or classified as worker, compared to 63% males. Among the working age group of 15-64 years, these percentages are 21% and 84%, respectively. Females are twice as likely to be illiterate than males and less likely to achieve any of the education levels. Females are 30% less likely than males to be never married, and about five times more likely to be separated or widowed. Among females, only 15% reported participating in any employment-related activity on the survey day compared to 63% males. The participation of females is much greater in household activities, with 25% participating in caregiving activities and 80% in domestic activities. Among males, these participation rates are 13% and 23%, respectively. Overall, 86% of the male respondents reported making at least one trip, compared to only 47.3% females. The average trip rate for females (1.32) is less than half the average trip rate for males (2.93), including all respondents.

**Figure 1a** presents percentage of respondents who reported making at least one trip on the survey day by age groups and gender. The level of mobility decreases sharply for females from their adolescence. For males, the only major reduction in their immobility is among older adults (64+ years). **Figure 1b** presents percentage individuals reporting their usual status as attending educational institution and those who reported being employed. The numbers presented here are running average over 3 age years for a smooth trend. Raw age-specific percentages are presented in **SI Table S3**. There are no significant gender differences in the education participation at any age. However, there is a wide gap in the work participation. Among males, work participation peaks around early 30s and remain so till late 40s. Among females, their work participation peaks around early 40s and is only for few years.

**Figure 2** presents two sets of gender-specific bivariate relationships. In the first set (figures a, b and c), relationships of trip rates are presented with age, household size, and monthly per capita household expenditure (MPCE), respectively. The second set (d, e and f) presents age-specific variation in participation time in different types of activities.

Number of daily trips for females fall steeply from the age of 6 years to late 20s, and then increases slightly to peak around 50s (see Figure 2a). The trip rates for males increase from 6 years to adolescence, drops in 20s, rises again to peak around 50s, and then fall again after age 50. At all ages, males have greater number of trips on average than females. Overall, magnitude of variation of trip rates is much greater among females than males. This trend with age can be explained by the variation in the participation of household-based activities, as shown in Figure 2d. The steep fall in trip rates for females in the childhood and adolescence correlates with a steep increase in household activities. Additionally, the increase in the trip rate after late 20s, correlates with a reduction in participation of household activities.

Trip rates for females fall steeply from living alone to living in a two-member household, following which there is further gradual reduction as household size increases (see Figure 2b). Among males, trip rates first increase as household size increases from one to three, and then reduce gradually thereafter. Again, the range of variation is much smaller among males than females. The relationship between standardised MPCE (a measure of income) and trip rate is much steeper for females than males (see Figure 2c). Among females, trip rates increase monotonically with MPCE. Among males, trip rates increase steeply from the lowest income level to half standard deviation increase in standardised MPCE, and then remain flat.

Figures 2d to 2f show a stark dichotomy in the activity participation of males and females. Across all age range, females participate in greater amount of domestic and caregiving activities than males, while males participate in greater employment activities than females. There is also a wide gap in the participation time. While females participate in more than 6 hours of domestic activity in their 30s, males participate in less than 30 minutes per day. Among males, participation time in domestic activities is almost never greater than an hour. Caregiving activities (for children and elderly) have greater variation across the age compared to domestic activities. The highest level of caregiving activity participation for females is around early 30s and is greater than 2 hours per day, while the highest for males is around mid-30s and is only as high as half an hour. The gender gap in participation time reverses for employment activities. While males participate in up to 7 hours of employment activity from the age of 30 to 60, female participate only for 2 hours. The fall in employment participation time starts from age 40 for females, while this fall occurs about two decades later for males.

### 4.1 Regression models

**Figure 3** presents the results of the first set of logistic regression models, without controlling for participation in different activity types. The results are presented as odds ratios with 95% confidence intervals. The results for both set of models are presented in the tabular format in the SI. The three variables that have the greatest effects in terms of magnitude are type of day (usual or not), age, and working status. The four variables viz. age, marital status, household size, and age of youngest household member, have opposite effects for females and males.

In an unusual day, both females and males are less likely to make a trip, though this likelihood is greater for males. Compared to 6-9 age group (reference), females in 10-14 age group are 30% **less** likely to make a trip, while males of the same age group are 18% **more** likely. For all other age groups, both females and males have lower likelihood of making the trip than the reference age, though the effect is much greater for females. For example, women aged 45-64 years are 62% less likely to make a trip and men of the same age are 33% less likely.

Married women are 40% **less** likely to make a trip than single women, while married men are about 9% **more** likely than single men. Compared to a single-member household, females in household size of 2 or greater are **less** likely to make a trip, and they are progressively less likely to with increasing household size. Among males, the effect is in opposite direction, and they are **more** likely to make a trip with household size of 2 or greater. For a household size of 5 or more, females are 20% **less** likely to make a trip than a single-member household, while males are 27% **more** likely. The relationship is steeper for females while it is largely flat for males.

Females in households with no infant are more likely to make a trip than those in households with an infant (reference). If the youngest member of the household is 6-to-14 years of age, females in those household have 45% greater likelihood to make a trip. Among males, this effect is only significant for the youngest age of 15 years or more, for which males are 11% **less** likely to make a trip, while females are 29% **more** likely.

For females and males, education attainment is associated with greater likelihood to make a trip compared to illiteracy. At each education level, effect size is smaller for females than males. For example, a female with graduation or a higher degree is 17% more likely than an illiterate to make a trip. In comparison, a male with the same education is 57% more likely. For both females and males, higher per capita expenditure is associated with greater likelihood of mobility. However, the effect is much greater and the relationship is much steeper among females compared to males. Females in the highest income quintile are 63% more likely to make a trip than those in the lowest quintile. Males, on the other hand, are 23% more likely.

Compared to those not working (as reference), females and males who are studying or are employed, have much greater likelihood of making a trip. Female workers are more than 9 times likely to make a trip than those not working. Females who are self-employed or are studying are 2 to 5 times more likely. Among males, the magnitude of effects for the working status are smaller, but in the same direction. For example, working males are about 6 times more likely to make a trip, and self-employed or studying males have similar odds as females.

Mechanical washing of clothes is associated with greater likelihood of making a trip for both females and males (8-11%) compared to manual washing. Outsourcing of washing does not have a significant effect for either group. Mechanical sweeping of floors compared to manual sweeping (reference) does not have a significant effect for either gender. However, outsourcing of sweeping is associated with 28% greater likelihood of making a trip for females and 25% greater likelihood for males.

Participation time in domestic activities have opposite effects on females and males. Females participating in domestic activities for less than 2 hours, and for 2 to 5 hours every day, are **less** likely to make a trip than those who participate in no domestic activity. Males participating in domestic activity for up to 5 hours are **more** likely to make a trip. For participation time in domestic activities greater than 5 hours, both females and males are less likely to make a trip (58% and 28%, respectively).

Participation in caregiving and employment activities have the same direction of association for males and females with almost the same effect size. With caregiving activities, there is an inverted U-shaped relationship with the likelihood of making a trip. Less than an hour per day of caregiving is associated with greater likelihood (14-22%) of making a trip compared to no caregiving. However, greater than an hour per day of caregiving is associated with 15% lower likelihood of making a trip.

For participation in employment activities, the effect size is greater for males up to 6 hours of participation, and greater for females for longer than 6 hours. Employment participation of 6 hours or less compared to none is associated with 75% greater likelihood of making a trip among females, and 125% greater likelihood among males. For employment participation of greater than 8 hours, females are 140% more likely to make a trip, while males are 80% more likely.

The inclusion of activity-specific participation time in the model results in attenuation of some effects while strengthening of others, but the directions of effects remained the same. The greatest effect is for working status. For example, the odds ratio for a female worker reduces from 9.5 to 4.4 for females, and 5.6 to 3.4 for males. Similarly, the odds ratio of self-employment is also halved or even smaller. The effects of age groups are slightly weakened for both females and males and the effect of marriage is weakened for females but is almost the same for males.

### 4.2   Work-related mobility by industry type

**Table 2** shows the industrial classification of those who reported themselves as employed. Also shown for each industry type, the percentage of those who reported they work in a household enterprise and the percentage who reported making a trip to work. The table is sorted by the largest industrial sector employing women on top. The three largest sectors of employment for women are manufacturing, education, and wholesale/retail trade, employing 51% of working women. For men also, manufacturing and wholesale/retail trade are among the three largest sectors along with construction, employing 56% of working men. Men and women are equally likely to be employed in manufacturing. The largest differences are in education, health, and households as employers where women are four to seven times more likely to be employed than men. While, in construction and transportation, women are 60-80% less likely to be employed than men.

The table also shows that, for a given industrial sector, men are often more likely to report a work trip than women. This difference is the greatest in the manufacturing sector, where only 33% of women employed in this sector reported travelling to work compared to 74% of men. This can be explained by a much greater proportion of women in manufacturing working in household enterprises compared to men (56% vs 35%). In wholesale/retail trade, both men and women are equally highly likely to be employed in household enterprise (70%), but men are still more likely to report a working trip than women.

### 4.3   Trip purpose

**Table 3** shows the percentage distribution of trip purpose for each age and gender group. The distributions by trip purpose are most similar up to age 9 for the two gender groups. The divergence in the distributions starts at 10-15 age group—with females reporting greater share of education trips but smaller share of recreation trips than males. Starting age 16, when individuals can legally work, in addition to recreational and education trips, the two groups diverge in work trips also, which males are three times as likely to make as females. In the 20-24 age group, there is additional divergence in the shopping or errand trips, which females are more likely to make than males. From 25-44 onwards, education trips are only a small minority for both groups. For males, work trips have a large majority of share, and for females, work, shopping, and social trips have almost similar share.

## 5   DISCUSSION
### 5.1   Statement of principal findings

I used a nationally representative time-use survey of India and analysed this data for urban residents. Two outcomes have been reported—average trip rate and mobility rate, where the latter is defined as the percentage of individuals who made at least one trip on the reporting day. I used exploratory analyses and gender-stratified individual-level logistic regression with mobility (making at least one trip on the reporting day) as the binary outcome.

The mobility rates for females and males are 47% and 86%, respectively. In other words, 53% of females reported not going out of home on the reporting day compared to only 14% males. The average trip rate is 1.3 among females and 2.9 among males. The gender gap in mobility is strongly age dependent. Mobility rate of females decrease steeply from adolescence to young adulthood, beyond which this rate is relatively stable at the low level. There is a slight increase in mobility in the middle age group. No such variation is observed in the mobility among males.

There is a clear dichotomy in gender roles with women participating largely in unpaid in-house activities while men in activities outside home. For example, women aged 25-44 years spend an average of 8.5 hours per day on domestic or caregiving activities (**SI Table A2**). In contrast, men in the same age group spend less than an hour on these activities. The participation level in out-of-home mobility reverses for the two gender groups. Only 38% women of this age group reported going out of home compared to 88% men. The impact of gender roles on mobility is reflected in the effects of household composition as found in the regression model. For example, being married or living with one person or more is associated with *reduced* mobility of women but *increased* mobility of men. Having an infant in the house is associated with reduced mobility of women but has almost no effect on mobility of men. The results clearly show that gender is a strong effect modifier, and highlight the importance of gender-stratified analyses i.e. not treating gender as a binary confounder.

It is interesting to note that the effect of age on female mobility remains strong even after controlling for multiple variables. The age-specific behaviours such as education level, work status, marital status, or even the participation time in different in-house activities do not weaken the age effect. This is likely the effect of social norms that have not been not been control for in the analysis.

This is the first study that reports gender differences in mobility in India at the national level. By using a large dataset of more than 170,000 individuals, our study provides strong evidence of gender inequality in terms of daily mobility in India. While there was evidence for such differences in the literature, this was contributed by small-scale studies in various urban areas of India. Additionally, none of the studies analysed mobility rate as primary outcome and its various determinants.

## 5.2    Comparison with other settings

Madre et al. (2007) reported gender-specific level of immobility in one-day travel diary surveys across 15 European countries. In all the countries except one immobility was *greater* among males than females. In our study, we found a reverse relationship. Overall, the average share of immobile was 8-12% across the countries reported by Madre et al. (2007). The upper value is similar to 14% among males in our study, but much smaller than females in our study. For London, Schmöcker et al. (2005) reported no gender differences in total number of trips per individual. In France, women made *greater* number of trips than men (Havet et al., 2021). For 18 cities across the world from Australia, Europe, Latin America, Sub-Saharan Africa, and North America, Goel et al. (2022) reported than on average 76% of females reported making a trip compared to 79% males. This indicates that the large gender disparity in mobility rate in India is an outlying phenomenon not commonly observed in most parts of the world.

Literature indicates that this large gender gap in mobility may be common in South Asian and Arab settings. For Pakistan, Adeel et al. (2013) used the time use survey and reported that only 45% females reported making a trip compared to 96% males. For females, this is very close to the 47% figure reported in our study. For males, Indians reported a lower mobility rate (88%) than Pakistanis. Another example is for Arab population in Israel, where a greater proportion of women (22%) reported not travelling at all compared to men (6%) (Elias et al., 2015). Countries in South Asia, Middle East and North Africa are among the most gender unequal societies in the world, with lower work participation rate among women compared to the rest of the world (Kabeer et al., 2019). The similarity across India, Pakistan, and Arab population is therefore in line with the similarity of these settings in gender inequality.

## 5.3    Future research implications

As highlighted in the introduction section, there is plenty of research in India that focusses on the low work participation rate of women. These research, mostly conducted by economists, often limit the analysis to 15 years or older, which is the age cut-off for working age population. However, my research findings point to the need for a life-course approach i.e. analysing determinants of mobility throughout the life of individuals. The results show that the mobility of females start reducing from adolescence and stabilises at a low level starting from young adulthood. That means the social norms that restrict women from working outside home, or going out of home at all, start their effect early in the childhood. The investigation of such research questions would need panel or longitudinal datasets. The results show that women who are not employed are also least likely to make *any trip* outside home. Only 30% such women reported going out of home at least once, compared to 81% women who are pursuing education or working. In other words, it is not just that some women are not going out to work, but many of them are not going out of home at all. Thus, the issue of women working outside home should be linked with a broader issue of women's mobility outside home.

**Table 1: Gender-specific characteristics and trip rates**

| Variables | Categories | Sample distribution | | Trip rates | | Mobility (%) | |
|---|---|---|---|---|---|---|---|
| | | female (N = 84207) | male (N = 88914) | female | male | female | male |
| Age category | All | | | 1.32 | 2.9 | 47 | 86 |
| | 6-9 years | 4.3 | 4.7 | 2.88 | 3.02 | 88 | 89 |
| | 10-15 years | 8.9 | 10.1 | 2.53 | 3.12 | 84 | 91 |
| | 16-19 years | 6.7 | 7.8 | 1.76 | 2.94 | 66 | 88 |
| | 20-24 years | 9.9 | 10.0 | 1.22 | 2.75 | 47 | 86 |
| | 25-44 years | 38.9 | 37.6 | 1.01 | 2.93 | 38 | 88 |
| | 45-64 years | 24.1 | 22.7 | 1.14 | 3.04 | 41 | 85 |
| | 64+ years | 7.1 | 7.1 | 0.88 | 2.44 | 31 | 66 |
| Work status | Not working | 61.9 | 12.4 | 0.79 | 2.34 | 29 | 65 |
| | Education | 21.9 | 24.6 | 2.36 | 3.01 | 81 | 90 |
| | Self employed | 5.2 | 24.7 | 1.48 | 3.10 | 49 | 84 |
| | Worker | 10.9 | 38.3 | 2.16 | 2.96 | 81 | 92 |
| Education | Illiterate | 16 | 7.7 | 1.00 | 2.71 | 36 | 79 |
| | Below_10th | 51.5 | 53.2 | 1.43 | 2.99 | 50 | 86 |
| | 11th_12th | 12.1 | 13.3 | 1.25 | 2.91 | 47 | 87 |
| | Diploma | 2.8 | 4.5 | 1.31 | 2.87 | 51 | 87 |
| | Graduate+ | 17.6 | 21.3 | 1.34 | 2.89 | 50 | 87 |
| Marital status | Never married | 28.1 | 40.3 | 2.17 | 2.91 | 75 | 88 |
| | Married | 60.4 | 57.1 | 0.94 | 2.97 | 35 | 86 |
| | Widowed/separated | 11.5 | 2.6 | 1.23 | 2.26 | 43 | 66 |
| Household expenditure per person (quintiles) | 1 | 24.8 | 23.6 | 1.30 | 2.80 | 44 | 84 |
| | 2 | 22.2 | 21.9 | 1.23 | 2.88 | 45 | 86 |
| | 3 | 20.3 | 20.5 | 1.25 | 2.96 | 47 | 86 |
| | 4 | 17.9 | 18.5 | 1.31 | 2.98 | 50 | 87 |
| | 5 | 14.7 | 15.4 | 1.38 | 2.92 | 54 | 87 |
| Household size | 1 | 3.0 | 5.0 | 1.90 | 2.81 | 61 | 84 |
| | 2 | 11.2 | 10.1 | 1.25 | 2.90 | 45 | 83 |
| | 3-4 | 46.2 | 47.2 | 1.33 | 2.99 | 48 | 88 |
| | 5+ | 39.6 | 37.6 | 1.29 | 2.88 | 46 | 85 |
| Age of youngest member | 0-1 years | 8.1 | 7.1 | 0.88 | 2.92 | 32 | 86 |
| | 2-5 years | 23 | 20.7 | 1.18 | 2.91 | 41 | 87 |
| | 6-14 years | 24.6 | 23.8 | 1.67 | 3.08 | 58 | 88 |
| | 15+ years | 44.3 | 48.4 | 1.27 | 2.86 | 47 | 84 |
| 5-year+ needing special care | no | 58.6 | 57.9 | 1.26 | 2.81 | 46 | 86 |
| | yes | 3.9 | 3.7 | 1.10 | 2.62 | 40 | 80 |
| Washing clothes | Manual | 75.3 | 75.7 | 1.29 | 2.93 | 46 | 86 |
| | Mechanical | 21.7 | 21.2 | 1.37 | 2.93 | 50 | 87 |
| | Outsourced | 3.0 | 3.1 | 1.62 | 2.92 | 56 | 84 |
| Sweeping floor | Manual | 89.8 | 89.7 | 1.3 | 2.94 | 47 | 86 |
| | Mechanical | 4.0 | 4.2 | 1.29 | 2.61 | 48 | 84 |
| | Outsourced | 6.2 | 6.1 | 1.63 | 2.95 | 57 | 86 |
| Survey response | Self | 62.8 | 50.2 | 1.19 | 2.95 | 43 | 84 |
| | Other household member | 36.6 | 49.1 | 1.53 | 2.91 | 55 | 88 |
| | Non-household member | 0.5 | 0.7 | 1.47 | 2.78 | 55 | 87 |
| Type of day for the respondent | Normal | 91.3 | 89.5 | 1.28 | 2.92 | 47 | 87 |
| | Other | 8.7 | 10.5 | 1.70 | 2.98 | 54 | 78 |

| Variables | Categories | Sample distribution | | Trip rates | | Mobility (%) | |
|---|---|---|---|---|---|---|---|
| | | female (N = 84207) | male (N = 88914) | female | male | female | male |
| Time spent in domestic activities | No activity | 20.4 | 76.3 | 2.25 | 2.78 | 76 | 86 |
| | 0-2 hours | 8.2 | 14.5 | 1.93 | 3.68 | 68 | 91 |
| | 2-5 hours | 18.5 | 7.6 | 1.43 | 3.12 | 52 | 85 |
| | 5+ hours | 52.8 | 1.5 | 0.83 | 2.35 | 31 | 66 |
| Time spent in care activities | No activity | 73.5 | 86.8 | 1.47 | 2.92 | 52 | 86 |
| | 0-1 hour | 6.1 | 6.6 | 1.18 | 3.12 | 44 | 89 |
| | 1+ hours | 20.4 | 6.6 | 0.82 | 2.83 | 30 | 84 |
| Time spent in employment activities | No activity | 84.8 | 42.7 | 1.21 | 2.8 | 43 | 80 |
| | <6 hours | 6.4 | 8.0 | 1.71 | 3.45 | 61 | 90 |
| | 6-8 hours | 5.0 | 17.3 | 2.15 | 3.21 | 79 | 92 |
| | 8+ hours | 3.8 | 32.1 | 2.09 | 2.82 | 81 | 89 |

**Table 2: Gender differences in employment characteristics and travel to work by industrial sectors**

| Industry | Employment distribution | | Percentage working in household enterprise | | Percentage reporting work trip | |
|---|---|---|---|---|---|---|
| | Female | Male | Female | Male | Female | Male |
| Manufacturing | 21% | 21% | 56% | 35% | 33% | 74% |
| Education | 18% | 4% | 11% | 13% | 65% | 73% |
| Wholesale and retail trade | 12% | 20% | 71% | 70% | 63% | 82% |
| Human health and social work activities | 8% | 2% | 16% | 29% | 72% | 74% |
| Activities of households as employers | 7% | 1% | 0% | 0% | 63% | 53% |
| Other service activities | 6% | 4% | 48% | 64% | 54% | 65% |
| Construction | 6% | 15% | 5% | 14% | 64% | 79% |
| Administrative and support service activities | 4% | 5% | 17% | 21% | 68% | 76% |
| Information and communication | 4% | 4% | 8% | 17% | 79% | 80% |
| Public administration and defence | 4% | 4% | 0% | 0% | 78% | 80% |
| Accommodation and Food service activities | 3% | 3% | 51% | 52% | 52% | 71% |
| Financial and insurance activities | 3% | 3% | 8% | 18% | 77% | 80% |
| Professional, scientific and technical activities | 2% | 2% | 21% | 34% | 73% | 77% |
| Transportation and storage | 1% | 8% | 20% | 48% | 71% | 61% |
| Arts, entertainment and recreation | 1% | 1% | 39% | 44% | 47% | 68% |
| Water supply; sewerage, waste management | 1% | 1% | 19% | 32% | 58% | 71% |
| Electricity, gas supply | 0% | 1% | 5% | 21% | 78% | 83% |
| Activities of extraterritorial organizations | 0% | 0% | 0% | 0% | 71% | 62% |
| Real estate activities | 0% | 1% | 17% | 80% | 83% | 75% |

**Table 3: Age and gender specific trip purpose distribution (each row sums to 100%)**

| Age | Gender | Education | Work | Shopping/ | Recreation | Social | Other |
|---|---|---|---|---|---|---|---|
| 6-9 | Female | 50.9 | 1.5 | 3.3 | 39.3 | 3.9 | 1 |
| | Male | 49.5 | 1.8 | 3.2 | 40.7 | 3.8 | 1.1 |
| 10-15 | Female | 60.3 | 2.2 | 4.1 | 26.5 | 6 | 0.9 |
| | Male | 49.2 | 2.9 | 3.5 | 37.2 | 5.5 | 1.6 |
| 16-19 | Female | 63.2 | 5.6 | 8.4 | 9.1 | 12.1 | 1.5 |
| | Male | 38.5 | 15 | 8.1 | 22.9 | 13.3 | 2.2 |
| 20-24 | Female | 34.7 | 19.6 | 17.4 | 6.6 | 18.5 | 3.3 |
| | Male | 18.7 | 38.6 | 10.4 | 13.1 | 16.6 | 2.5 |
| 25-44 | Female | 2.5 | 32.8 | 30.4 | 5.9 | 20.9 | 7.6 |
| | Male | 1.3 | 62.8 | 13.5 | 6.4 | 13 | 3.1 |
| 45-64 | Female | 0.4 | 26.7 | 30.4 | 11.4 | 26.6 | 4.4 |
| | Male | 0.2 | 54.7 | 17.1 | 9.5 | 14.8 | 3.7 |
| 64+ | Female | 0.3 | 11.1 | 26 | 17.6 | 38.9 | 6.2 |
| | Male | 0.1 | 23.8 | 23.7 | 21.2 | 25.8 | 5.4 |

19 Aug 2022

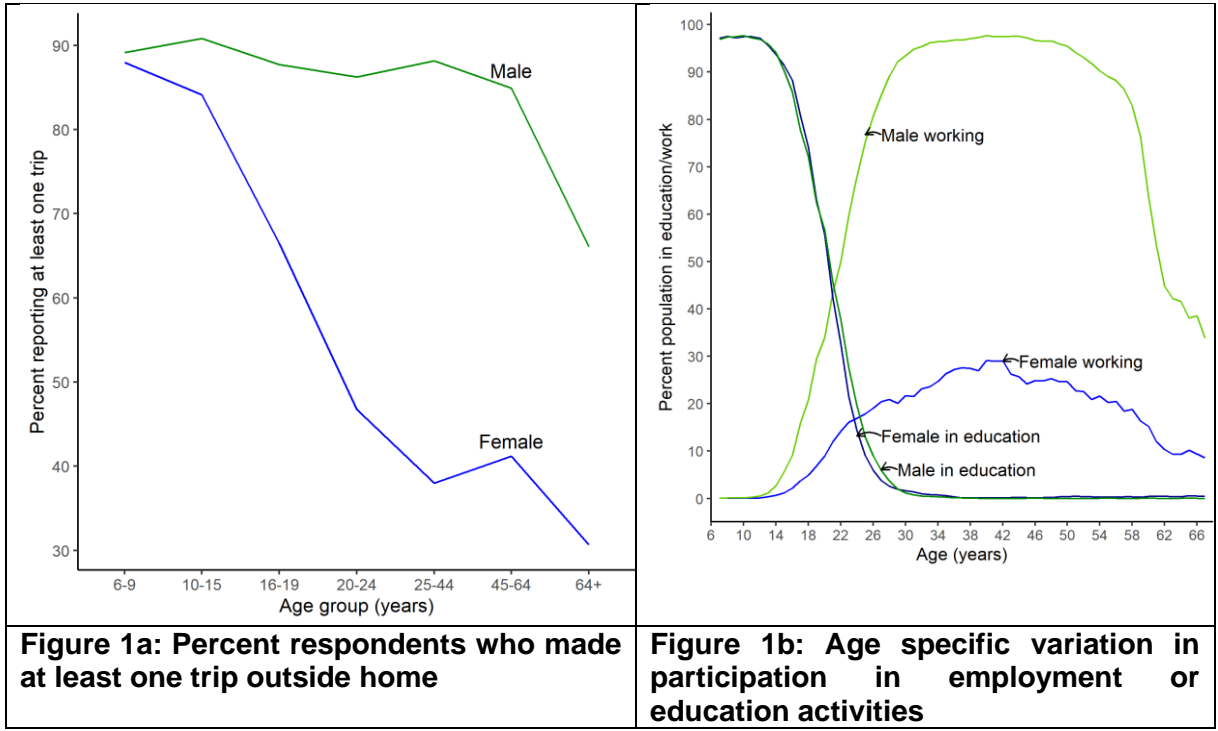

Figure 1a: Percent respondents who made at least one trip outside home

Figure 1b: Age specific variation in participation in employment or education activities

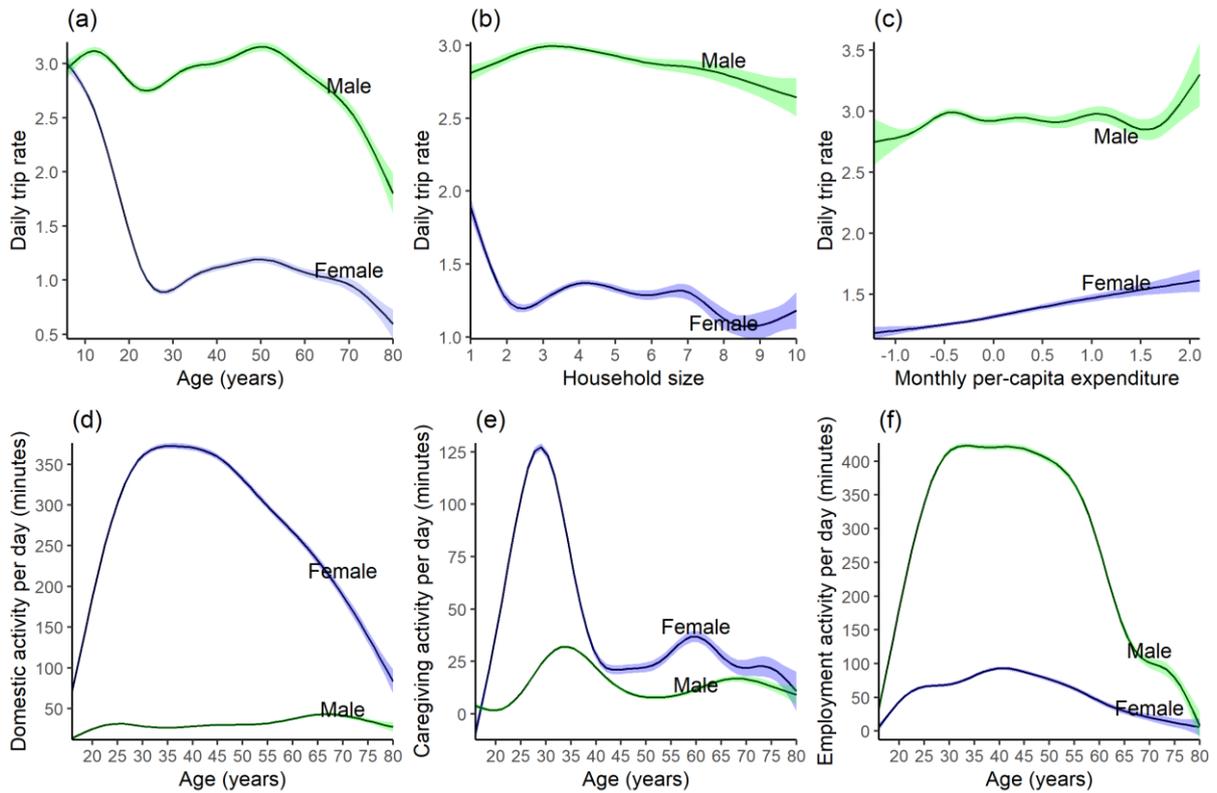

Figure 2: GAMM models

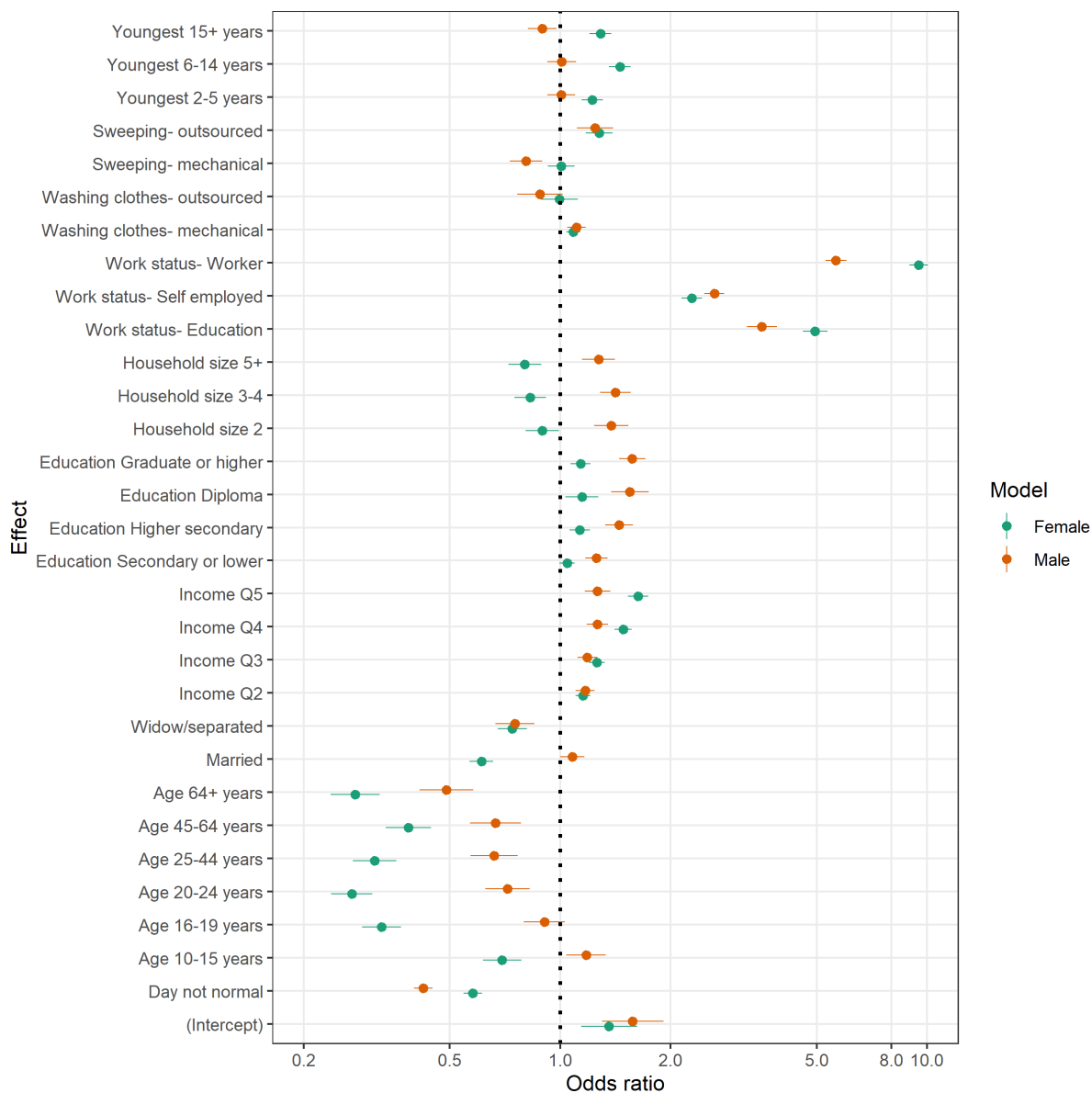

Figure 3: Gender specific regression model with out-of-home mobility as binary outcome [Reference categories: Day normal, Age 06-09 years, Never married, Income Q1, Education illiterate, Household size 1, Work status- not working, Washing clothes- manual, Sweeping- manual, Youngest 0-1 years.

19 Aug 2022

19 Aug 2022

19 Aug 2022

**Supplementary material**
**Table A1: Activities that were reported to occur outside home and did not have accompanying travel activity**

| Activity code | Definition |
|---|---|
| 131 | Vending and trading of goods in household enterprises |
| 371 | Shopping for/purchasing of goods and related activities |
| 611 | School/university attendance: For this activity, there were two instances for a given individual indicating trip to school and back. One trip was assigned to each instance that covers the round trip. It was only assigned for those individuals that do not make any learning-related trip with the activity code of 640: Travelling time related to learning |
| 712 | Socializing/getting together/ gathering activities |
| 742 | Participating in collective religious practice |
| 823 | Playing games and other pastime activities |
| 829 | Other activities related to cultural participation, hobbies, Games |
| 831 | Participating in sports |
| 832 | Exercising |
| 950 | Travelling time related to self-care and maintenance activities |
| 442, 443, 444 | Accompanying child, dependent and other adult: Assigned 2 trips wherever one trip was mentioned to cover the round trip. |

**Table A2: Activity participation per capita by gender and age groups (minutes per capita)**

| Gender | Activity type | Age group | | | | | | |
|---|---|---|---|---|---|---|---|---|
| | | 06-09 | 10-15 | 16-19 | 20-24 | 25-44 | 45-64 | 64+ |
| Female | Employment and related activities | 2 | 5 | 24 | 67 | 92 | 79 | 24 |
| Female | Production of goods for own final use | 0 | 1 | 4 | 5 | 8 | 10 | 7 |
| Female | Unpaid domestic services | 5 | 34 | 131 | 277 | 426 | 367 | 192 |
| Female | Unpaid caregiving services | 8 | 6 | 9 | 63 | 86 | 27 | 23 |
| Female | Unpaid volunteer, trainee and other unpaid work | 1 | 2 | 3 | 3 | 2 | 3 | 2 |
| Female | Learning | 385 | 426 | 334 | 140 | 9 | 1 | 1 |
| Female | Socialising, community and religious participation | 105 | 131 | 176 | 182 | 178 | 219 | 253 |
| Female | Culture, leisure, mass-media and sports practices | 263 | 222 | 198 | 195 | 188 | 223 | 288 |
| Female | Self-care and maintenance | 226 | 229 | 230 | 230 | 231 | 240 | 257 |
| Male | Employment and related activities | 3 | 11 | 107 | 287 | 480 | 417 | 126 |
| Male | Production of goods for own final use | 0 | 1 | 2 | 4 | 7 | 10 | 10 |
| Male | Unpaid domestic services | 3 | 7 | 20 | 32 | 33 | 39 | 44 |
| Male | Unpaid caregiving services | 5 | 5 | 3 | 4 | 24 | 11 | 15 |
| Male | Unpaid volunteer, trainee and other unpaid work | 0 | 1 | 3 | 3 | 4 | 4 | 5 |
| Male | Learning | 397 | 425 | 332 | 168 | 14 | 2 | 1 |
| Male | Socialising, community and religious participation | 105 | 124 | 178 | 195 | 193 | 215 | 276 |
| Male | Culture, leisure, mass-media and sports practices | 267 | 249 | 223 | 195 | 163 | 204 | 330 |
| Male | Self-care and maintenance | 225 | 226 | 230 | 231 | 249 | 258 | 275 |

19 Aug 2022

**Table A3: Gender-stratified logistic regression model with out-of-home mobility as binary outcome [mobility is whether the individual reported going out of home at least once on the reporting day]**

|  | Female | | Male | |
|---|---|---|---|---|
| Variable | Coefficients | p-value | Coefficients | p-value |
| (Intercept) | 1.36 (1.14, 1.62) | <0.001 | 1.58 (1.3, 1.91) | <0.001 |
| Day normal | 1 | | 1 | |
| Day not normal | 0.58 (0.54, 0.61) | <0.001 | 0.42 (0.4, 0.45) | <0.001 |
| Age 06-09 years | 1 | | 1 | |
| Age 10-15 years | 0.69 (0.62, 0.78) | <0.001 | 1.18 (1.04, 1.33) | <0.01 |
| Age 16-19 years | 0.33 (0.29, 0.37) | <0.001 | 0.91 (0.8, 1.03) | 0.137 |
| Age 20-24 years | 0.27 (0.24, 0.31) | <0.001 | 0.72 (0.63, 0.83) | <0.001 |
| Age 25-44 years | 0.31 (0.27, 0.36) | <0.001 | 0.66 (0.57, 0.77) | <0.001 |
| Age 45-64 years | 0.39 (0.33, 0.44) | <0.001 | 0.67 (0.57, 0.78) | <0.001 |
| Age 64+ years | 0.28 (0.24, 0.32) | <0.001 | 0.49 (0.41, 0.58) | <0.001 |
| Never married | 1 | | 1 | |
| Married | 0.61 (0.57, 0.66) | <0.001 | 1.08 (1, 1.17) | <0.1 |
| Widow/separated | 0.74 (0.68, 0.81) | <0.001 | 0.75 (0.67, 0.85) | <0.001 |
| Income Q1 | 1 | | 1 | |
| Income Q2 | 1.15 (1.1, 1.21) | <0.001 | 1.17 (1.1, 1.24) | <0.001 |
| Income Q3 | 1.26 (1.2, 1.32) | <0.001 | 1.19 (1.12, 1.26) | <0.001 |
| Income Q4 | 1.48 (1.41, 1.57) | <0.001 | 1.26 (1.18, 1.35) | <0.001 |
| Income Q5 | 1.63 (1.53, 1.74) | <0.001 | 1.26 (1.17, 1.37) | <0.001 |
| Education Illiterate | 1 | | 1 | |
| Education Secondary or lower | 1.04 (1, 1.1) | <0.1 | 1.26 (1.17, 1.35) | <0.001 |
| Education Higher secondary | 1.13 (1.06, 1.2) | <0.001 | 1.45 (1.33, 1.58) | <0.001 |
| Education Diploma | 1.15 (1.03, 1.27) | <0.01 | 1.55 (1.38, 1.75) | <0.001 |
| Education Graduate or higher | 1.14 (1.07, 1.21) | <0.001 | 1.57 (1.45, 1.71) | <0.001 |
| Household size 1 | 1 | | 1 | |
| Household size 2 | 0.89 (0.8, 0.99) | <0.05 | 1.38 (1.24, 1.53) | <0.001 |
| Household size 3-4 | 0.83 (0.75, 0.91) | <0.001 | 1.41 (1.29, 1.56) | <0.001 |
| Household size 5+ | 0.8 (0.72, 0.89) | <0.001 | 1.27 (1.15, 1.41) | <0.001 |
| Work status- Not working | 1 | | 1 | |
| Work status- Education | 4.95 (4.58, 5.35) | <0.001 | 3.55 (3.23, 3.9) | <0.001 |
| Work status- Self employed | 2.28 (2.14, 2.43) | <0.001 | 2.63 (2.47, 2.8) | <0.001 |
| Work status- Worker | 9.49 (8.96, 10.05) | <0.001 | 5.64 (5.29, 6.02) | <0.001 |
| Washing clothes- manual | 1 | | 1 | |
| Washing clothes- mechanical | 1.09 (1.04, 1.14) | <0.001 | 1.11 (1.05, 1.17) | <0.001 |
| Washing clothes- outsourced | 1 (0.89, 1.12) | 0.94 | 0.88 (0.76, 1.02) | <0.1 |
| Sweeping- manual | 1 | | 1 | |
| Sweeping- mechanical | 1.01 (0.93, 1.09) | 0.888 | 0.81 (0.73, 0.89) | <0.001 |
| Sweeping- outsourced | 1.28 (1.17, 1.39) | <0.001 | 1.25 (1.11, 1.4) | <0.001 |
| Youngest 0-1 year | 1 | | 1 | |
| Youngest 2-5 years | 1.22 (1.14, 1.31) | <0.001 | 1.01 (0.92, 1.1) | 0.878 |
| Youngest 6-14 years | 1.45 (1.36, 1.56) | <0.001 | 1.01 (0.92, 1.1) | 0.837 |
| Youngest 15+ years | 1.29 (1.2, 1.38) | <0.001 | 0.89 (0.82, 0.98) | <0.05 |

19 Aug 2022

**Table A4: Gender-stratified logistic regression model with out-of-home mobility as binary outcome controlling for activity participation time[mobility is whether the individual reported going out of home at least once on the reporting day]**

|  | Female |  | Male |  |
|---|---|---|---|---|
| Variable | Coefficients | p-value | Variable | Coefficients |
| (Intercept) | 1.89 (1.58, 2.26) | <0.001 | 1.34 (1.1, 1.64) | <0.01 |
| Day normal | 1 |  | 1 |  |
| Day not normal | 0.66 (0.63, 0.7) | <0.001 | 0.52 (0.48, 0.55) | <0.001 |
| Age 06-09 years | 1 |  | 1 |  |
| Age 10-15 years | 0.75 (0.66, 0.85) | <0.001 | 1.16 (1.02, 1.31) | <0.05 |
| Age 16-19 years | 0.4 (0.36, 0.46) | <0.001 | 0.84 (0.74, 0.96) | <0.05 |
| Age 20-24 years | 0.35 (0.3, 0.39) | <0.001 | 0.65 (0.56, 0.74) | <0.001 |
| Age 25-44 years | 0.41 (0.35, 0.47) | <0.001 | 0.59 (0.51, 0.69) | <0.001 |
| Age 45-64 years | 0.46 (0.39, 0.53) | <0.001 | 0.6 (0.51, 0.7) | <0.001 |
| Age 64+ years | 0.26 (0.22, 0.31) | <0.001 | 0.46 (0.38, 0.54) | <0.001 |
| Never married | 1 |  | 1 |  |
| Married | 0.88 (0.82, 0.96) | <0.01 | 1.06 (0.98, 1.14) | 0.156 |
| Widow/separated | 0.88 (0.8, 0.97) | <0.05 | 0.78 (0.69, 0.88) | <0.001 |
| Income Q1 | 1 |  | 1 |  |
| Income Q2 | 1.15 (1.09, 1.2) | <0.001 | 1.17 (1.1, 1.24) | <0.001 |
| Income Q3 | 1.24 (1.18, 1.31) | <0.001 | 1.19 (1.12, 1.27) | <0.001 |
| Income Q4 | 1.47 (1.39, 1.55) | <0.001 | 1.26 (1.18, 1.35) | <0.001 |
| Income Q5 | 1.61 (1.51, 1.71) | <0.001 | 1.29 (1.19, 1.39) | <0.001 |
| Illiterate | 1 |  | 1 |  |
| Education Secondary or lower | 1.12 (1.07, 1.18) | <0.001 | 1.23 (1.14, 1.32) | <0.001 |
| Education Higher secondary | 1.22 (1.14, 1.3) | <0.001 | 1.44 (1.31, 1.57) | <0.001 |
| Education Diploma | 1.21 (1.09, 1.34) | <0.001 | 1.52 (1.35, 1.71) | <0.001 |
| Education Graduate or higher | 1.21 (1.14, 1.29) | <0.001 | 1.52 (1.4, 1.66) | <0.001 |
| Household size 1 | 1 |  | 1 |  |
| Household size 2 | 0.84 (0.76, 0.94) | <0.01 | 1.46 (1.31, 1.63) | <0.001 |
| Household size 3-4 | 0.74 (0.67, 0.82) | <0.001 | 1.55 (1.4, 1.71) | <0.001 |
| Household size 5+ | 0.67 (0.6, 0.74) | <0.001 | 1.43 (1.28, 1.59) | <0.001 |
| Work status- Not working | 1 |  | 1 |  |
| Work status- Education | 3.9 (3.6, 4.23) | <0.001 | 3.67 (3.34, 4.04) | <0.001 |
| Work status- Self employed | 1.16 (1.07, 1.27) | <0.001 | 1.59 (1.46, 1.72) | <0.001 |
| Work status- Worker | 4.37 (4.03, 4.73) | <0.001 | 3.39 (3.11, 3.69) | <0.001 |
| Washing clothes- manual | 1 |  | 1 |  |
| Washing clothes- mechanical | 1.08 (1.03, 1.13) | <0.001 | 1.12 (1.06, 1.19) | <0.001 |
| Washing clothes- outsourced | 0.92 (0.82, 1.04) | 0.181 | 0.9 (0.78, 1.04) | 0.137 |
| Sweeping- manual | 1 |  | 1 |  |
| Sweeping- mechanical | 0.98 (0.9, 1.07) | 0.692 | 0.81 (0.74, 0.9) | <0.001 |
| Sweeping- outsourced | 1.2 (1.1, 1.31) | <0.001 | 1.26 (1.13, 1.41) | <0.001 |
| Youngest 0-1 year | 1 |  | 1 |  |
| Youngest 2-5 years | 1.21 (1.13, 1.29) | <0.001 | 0.97 (0.89, 1.06) | 0.545 |
| Youngest 6-14 years | 1.41 (1.31, 1.52) | <0.001 | 0.97 (0.89, 1.07) | 0.584 |
| Youngest 15+ years | 1.23 (1.14, 1.33) | <0.001 | 0.86 (0.78, 0.94) | <0.01 |

19 Aug 2022

|  | **Female** |  | **Male** |  |
|---|---|---|---|---|
| **Variable** | **Coefficients** | p-value | **Variable** | **Coefficients** |
| Domestic activities None | 1 |  | 1 |  |
| Domestic activities 0-2 hours | 0.95 (0.88, 1.02) | 0.134 | 2.17 (2.03, 2.33) | <0.001 |
| Domestic activities 2-5 hours | 0.6 (0.57, 0.64) | <0.001 | 1.5 (1.39, 1.62) | <0.001 |
| Domestic activities 5+ hours | 0.42 (0.39, 0.44) | <0.001 | 0.72 (0.63, 0.81) | <0.001 |
| Caregiving activities None | 1 |  | 1 |  |
| Caregiving activities 0-1 hours | 1.14 (1.07, 1.22) | <0.001 | 1.22 (1.11, 1.35) | <0.001 |
| Caregiving activities 1+ hours | 0.85 (0.81, 0.9) | <0.001 | 0.85 (0.78, 0.93) | <0.001 |
| Employment activities None | 1 |  | 1 |  |
| Employment activities <6 hours | 1.75 (1.62, 1.9) | <0.001 | 2.25 (2.04, 2.48) | <0.001 |
| Employment activities 6-8 hours | 2.72 (2.45, 3.02) | <0.001 | 2.59 (2.37, 2.83) | <0.001 |
| Employment activities 8+ hours | 2.4 (2.13, 2.69) | <0.001 | 1.78 (1.65, 1.93) | <0.001 |